\documentclass[runningheads,a4paper]{llncs}

\usepackage{placeins}  
\usepackage{color}
\usepackage[usenames,dvipsnames]{xcolor}
\usepackage{url,graphicx,times}
\usepackage{tabularx}
\usepackage{multirow}
\usepackage{booktabs}
\usepackage[square,comma,numbers,sort&compress,sectionbib]{natbib}
\usepackage{enumerate}
\usepackage{amsmath}
\usepackage{hyperref}
\usepackage{amssymb}
\usepackage{graphicx}
\usepackage{hyperref}
\usepackage[ruled,vlined,linesnumbered]{algorithm2e}
\usepackage{cleveref}
\usepackage{enumitem}
\usepackage{floatrow}

\begin{document}

\mainmatter

\title{Generating High-Quality Query Suggestion Candidates for Task-Based Search}

\author{Heng Ding\inst{1,2}, Shuo Zhang\inst{2}, Dar\'{i}o Garigliotti\inst{2}, and Krisztian Balog\inst{2}}
\institute{Wuhan University, Wuhan, China \and
University of Stavanger, Stavanger, Norway\\
\email{
hengding@whu.edu.cn,
<firstname.lastname>@uis.no}
}

\maketitle

\begin{abstract}
We address the task of generating query suggestions for task-based search.  The current state of the art relies heavily on suggestions provided by a major search engine.  In this paper, we solve the task without reliance on search engines.
Specifically, we focus on the first step of a two-stage pipeline approach, which is dedicated to the generation of query suggestion candidates.  We present three methods for generating candidate suggestions and apply them on multiple information sources.  Using a purpose-built test collection, we find that these methods are able to generate high-quality suggestion candidates.
\end{abstract}

\section{Introduction}
\label{sec:intro}

Query suggestions, recommending a list of relevant queries to an initial user input, are an integral part of modern search engines~\citep{Ozertem:2012:LSM}.  Accordingly, this task has received considerable attention over the last decade~\citep{Mitra:2015:QAR,Bhatia:2011:QSA,Cai:2016:SQA}.    
Traditional approaches, however, do not consider the larger underlying task the user is trying to accomplish.  In this paper, we focus on generating query suggestions for supporting task-based search.  Specifically, we follow the problem definition of the \emph{task understanding} task from the TREC Tasks track: given an initial query, the system should return a ranked list of suggestions ``that represent the set of all tasks a user who submitted the query may be looking for''~\citep{Yilmaz:2015:OTT}.  Thus, the overall goal is to provide a complete coverage of aspects (subtasks) for an initial query, while avoiding redundancy.    

\begin{figure}[!t]
\centering
\includegraphics[width=8cm]{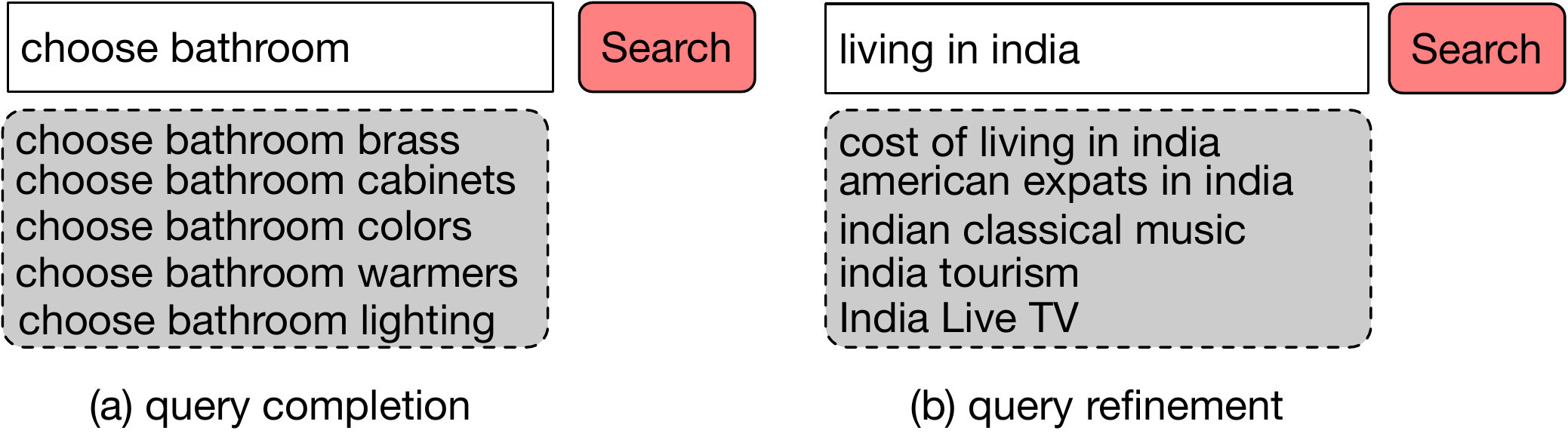}
\caption{Examples of query suggestions to support task-based search.}
\label{fig:ex}
\end{figure}

We envisage a user interface where task-based query suggestions are presented once the user has issued an initial query; see Fig.~\ref{fig:ex}.  
These query suggestions come in two flavors: \emph{query completions} and \emph{query refinements}. The difference is that the former are prefixed by the initial query, while the latter are not. 
It is an open question whether a unified method can produce suggestions in both flavors, or rather specialized models are required.  
The best published work on task-based query suggestions, that we know of, is by \citet{Garigliotti:2017:GQS}, who use a probabilistic generative model to combine keyphrase-based suggestions extracted from multiple information sources.  Nevertheless, they rely heavily on Google's query suggestion service.  
Thus, another main challenge in our work is to solve this task without relying on suggestions provided by a major web search engine (and possibly even without using a query log).

Following the pipeline architecture widely adopted in general query suggestion systems~\citep{Mitra:2015:QAR, Sordoni:2015:HRE}, we propose a two-step approach consisting of \emph{suggestion generation} and \emph{suggestion ranking} steps.   
In this paper, we focus exclusively on the first component.  Our aim is to generate sufficiently many high-quality query suggestion candidates.  The subsequent ranking step will then produce the final ordering of suggestions by reranking these candidates (and ensuring their diversity with respect to the possible subtasks).
The first research question we address is:  
\emph{Can existing query suggestion methods generate high-quality query suggestions for task-based search?}
Specifically, we employ the popular suffix~\citep{Mitra:2015:QAR}, neural language model~\citep{Park:2017:NLM}, and sequence-to-sequence~\citep{Sordoni:2015:HRE} approaches to generate candidate suggestions.
The second research question we ask is: \emph{What are useful information sources for each method?}
We are particularly interested in finding out how a task-oriented knowledge base (Know-How~\citep{Pareti:IKH:2016}) and a community question answering site (WikiAnswers~\citep{Fader:OQA:2014}) fare against using a query log (AOL~\citep{Pass:2006:PS}).
We find that the sequence-to-sequence approach performs best among the tested three methods. As for data sources, we observe that, as expected, the query log is the highest performing one among all. Nevertheless, the other two also provide valuable suggestion candidates that cannot be generated from the query log.  
Overall, we find that all method-source configurations contribute unique suggestions, and thus it is beneficial to combine them.  

\section{Query Suggestion Generation}
\label{sec:approach}

Given a task-related initial query, $q_0$, we aim to generate a list of query suggestions that cover all the possible subtasks related to the task the user is trying to achieve.  For a given suggestion $q$, let $P(q|q_0)$ denote the probability of that suggestion.
Below, we present three methods from the literature.  
The first two methods are specialized only in producing query completions, while the third one is able to handle both query completions and refinements.
Due to space constraints, only brief descriptions are given; we refer to the respective publications for further details.

\subsection{Popular Suffix Model}

The \emph{popular suffix model}~\citep{Mitra:2015:QAR} generates suggestions using frequently observed suffixes mined from a query log.  The method generates a suggestion $q$ by extending the input query $q_0$ with a popular suffix $s$, i.e., $q = q_0 \oplus s$, where $\oplus$ denotes the concatenation operator. 
The query likelihood is based on the popularity of suffix $s$: $P(q|q_0) = \mathit{pop}(s)$, where $\mathit{pop}(s)$ denotes the relative frequency of $s$ occurring in the query log.

\subsection{Neural Language Model}

Neural language models (NLMs) can be used for generating query suggestions~\citep{Park:2017:NLM}.  Suggestion $q$ is created by extending the input query $q_0$ character by character:  
$q = q_0 \oplus \mathbf{s} = (c_1,\dots,c_n,c_{n+1},\dots, c_{m})$,
where $c_1,\dots,c_n$ are the characters of $q_0$, and $\mathbf{s}=(c_{n+1},\dots,c_{m})$ are characters generated by the neural model.  Given a sequence of previous characters $\mathbf{c}=(c_1,\dots,c_{i})$, the model generates the next character ($i \geq n$) according to:
$P(c_{i+1}| \mathbf{c}) = \mathit{softmax}(h_{i})$,
where 
the hidden state vector at time $i$ is computed using $h_{i} = f(x_i, h_{i-1})$. Here, $f$ denotes a special unit, e.g., a long short-term memory (LSTM)~\cite{Hochreiter:1997:LSM}; $x_i$ is the vector representation of the $i$th character of suggestion $q$ and is taken to be $x_i = \sigma(c_i)$, where $\sigma$ denotes a mapping function from a character to its vector representation.
Finally, the query likelihood is estimated according to: 
\begin{equation*}
	\mathit{P(q|q_0)} = \prod_{j=n}^{m-1} {P(c_{j+1}| c_1,\dots,c_{j})} .
\end{equation*}
Our implementation uses a network of 512 hidden units.  We initialize the word-embedded vector with the pre-trained vector from Bing queries.\footnote{https://www.microsoft.com/en-us/download/details.aspx?id=52597}   
Beam search width is set to 30.

\subsection{Sequence to Sequence Model}

The sequence-to-sequence model (Seq2Seq)~\cite{Luong:2015:EAA} aims to directly model the conditional probability $P(w'_1, \dots , w'_m|w_1, \dots , w_n)$ of translating a source sequence $(w_1, \dots , w_n)$ to a target sequence $(w'_1, \dots , w'_m)$.  Thus, it lends itself naturally to implement our query suggestion task using Seq2Seq, by letting the initial query be the source sequence $q_0=(w_1, \dots , w_n)$ and the suggestion be the target sequence $q=(w'_1, \dots , w'_m)$. 
Typically, a Seq2Seq model consists of two main components: an encoder and a decoder.

The \emph{encoder} is a recurrent neural network (RNN) to compute a context vector representation $c$ for the original query $q_0$.
The hidden state vector of the encoder RNN at time $i \in [1 .. n]$ is given by: $h_{i} = f(w_{i}, h_{i-1})$, where ${w}_{i}$ is the $i$th word of the input query $q_0$, and $f$ is a special unit (LSTM).
The context vector representation is updated by $c = \phi(h_{1}, h_{2}, \dots, h_{n})$, where $\phi$ is an operation choosing the last state $h_{n}$.

The \emph{decoder} is another RNN to decompress the context vector $c$ and output the suggestion, $q=(w'_1, \dots , w'_m)$, through a conditional language model.
The hidden state vector of the decoder RNN at time $i \in [1 .. m]$ is given by $h_{i}^{'} = f'(h_{i-1}^{'}, w_{i-1}^{'}, c)$, where $w_{i}^{'}$ is the $i$th word of the suggestion $q$, and $f'$ is a special unit (LSTM).  
The language model is given by: $P(w_{i}^{'}| w_{1}^{'}, \dots , w_{i-1}^{'}, q_0) = g(h_{i}^{'}, w_{i-1}^{'}, c)$, where $g$ is a softmax classifier.
Finally, the Seq2Seq model estimates the suggestion likelihood according to: 
\begin{equation*}
	\mathit{P(q|q_0)} = \prod_{j=1}^{m-1} {P(w'_{j+1}| w'_1,\dots, w'_j, q_0)}~.
\end{equation*}
We use a bidirectional GRU unit with size 100 for encoder RNNs, and a GRU unit with size 200 for decoder RNNs. We employ an Adam optimizer with an initial learning rate of $10^{-4}$ and a dropout rate of 0.5.    
Beam search width is set to 100.

\section{Data Sources}
\label{sec:data}

We consider three independent information sources.
For the PopSuffix and NLM methods, we need a collection of short texts, $\mathcal{C}$.  For Seq2Seq, we need pairs of question-suggestion pairs, $\langle \mathcal{Q},\mathcal{S} \rangle$.

\begin{itemize}
	\item \emph{AOL query log}~\citep{Pass:2006:PS}: a large query log that includes queries along with anonymous user identity and timestamp.  We extract all queries from the log as $\mathcal{C}$.  We detect sessions (each session including multiple queries) using the same criterion as in~\citep{Sordoni:2015:HRE}. Then, we pair queries in the same session to obtain $\langle \mathcal{Q},\mathcal{S} \rangle$, where $\mathcal{Q}$ denotes a set of queries and $\mathcal{S}$ are suggestions paired against $\mathcal{Q}$.  In order to obtain more pairs, we extract all proper prefixes from the query, and pair them together. For example, given a query ``make a pancake'', we can construct two pairs $\langle$ ``make'', ``make a pancake'' $\rangle$ and $\langle$ ``make a'', ``make a pancake'' $\rangle$. This way, we end up with a total of 112K prefix-query pairs.  
	\item \emph{KnowHow}~\citep{Pareti:IKH:2016}: a knowledge base that consists of two and half million entries. Each triple $\langle s, p, o \rangle$ represents a fact about a human task, where the subject $s$ denotes a task (e.g., ``make a pancake'') and the object $o$ is a subtask (e.g., ``prepare the mix'').  We collect all subjects and objects as $\mathcal{C}$, and take all (142K) subject-object pairs to form $\langle \mathcal{Q},\mathcal{S} \rangle$. Additionally, prefixes from tasks (i.e., subjects and objects) are extracted to get more pairs, the same way as it is done for the AOL query log.  
	\item \emph{WikiAnswers}~\citep{Fader:OQA:2014}: a collection of questions scraped from WikiAnswers.com.\footnote{http://www.answers.com/Q/}  We detect task-related questions using a simple heuristic, namely, that a task-related question often starts with question constructions ``how do you'' or ``how to.''  These question constructions are removed from the questions to obtain $\mathcal{C}$ (e.g, ``how to change gmail password'' $\to$ ``change gmail password'').  This source can only be used for the PopSuffix and NLM methods, as it does not provide pairs for Seq2Seq.
\end{itemize}

\section{Experiment}

We design and conduct an experiment to answers our research questions.  
First, we collect a pool of candidate suggestions, by applying the  methods on different sources. Second, we collect annotations for each of these suggestions via crowdsourcing. Finally, we report and analyze the results.

\subsection{Pool Construction}
\label{sec:ex:pc}

We consider all queries (100 in total) from the TREC 2015 and 2016 Tasks tracks~\citep{Yilmaz:2015:OTT, Verma:2016:OTT}.  We combine the proposed methods (Sect.~\ref{sec:approach}) with various information sources (Sect.~\ref{sec:data}) for suggesting candidates.  We shall write $s$-$m$ to denote a particular configuration that uses method $m$ with source $s$.  
In addition, we also include the suggestions generated by (i) the keyphrase-based query suggestion system~\citep{Garigliotti:2017:GQS}, and (ii) the Google Query Suggestion Service (referred to as \emph{Google API} for short.
Two pools are constructed, one for query completions (QC) and one for query refinements (QR).
The pool depth is 20, that is, we consider (up to) the top-20 suggestions for each method.

\subsection{Crowdsourcing}
\label{sec:ex:cs}

Due to the fact that many of our candidate suggestions lack assessments in the TREC ground truth (and thus are considered irrelevant),
we obtain relevance assessments for all suggestions via crowdsourcing.  
Specifically, we use the Crowdflower platform, where 
a dynamic number of annotators (3-5) are asked to label each suggestion as relevant or non-relevant.
A suggestion is relevant if it targets for some more specific aspect of the original query, i.e., the suggestion must be related to the original query intent.  It does not have to be perfectly correct grammatically, as long as the intent behind is clearly understandable.
The final label is taken to be the majority vote among the assessors.
A total of 12,790 QC and 9,608 QR suggestions are annotated, at the total expense of 692\$.
Further details are available in the online appendix.\footnote{http://bit.ly/2BnSjhR}

\subsection{Results and Analysis}
\label{sec:ex:result}

Table~\ref{tbl:res_sg} presents a comparison of methods in terms of precision at cutoff points 10 and 20 (P@10 and P@20). 
Overall, we find that our methods can generate high-quality query suggestions for task-based search.  Our best numbers are comparable to that of the Google API.
It should, however, be noted that the Google API can only generate a limited number of suggestions for each query.  
The keyphrase-based method~\cite{Garigliotti:2017:GQS} is the highest performing of all; it is expected, as it combines multiple information sources, including suggestions from Google.

\begin{table*}[!t]
\centering
\caption{Precision for candidate suggestions generated by different configurations. For QC methods, we also report on recall (R) and cumulative recall (CR).} 
\begin{tabular}{ l p{1cm} p{1cm} p{1cm} p{1cm} p{1cm} p{1cm} } 
	\toprule
	Method & \multicolumn{4}{c}{QC} & \multicolumn{2}{c}{QR} \\
	& P@10 & P@20 & R & CR & P@10 & P@20 \\
	\midrule
	AOL-PopSuffix & 0.257 & 0.245 & 0.168 & 0.168 & - & - \\ 
	KnowHow-PopSuffix & 0.195 & 0.170 & 0.102 & 0.256 & - & - \\
	WikiAnswers-PopSuffix & 0.181 & 0.167 & 0.101 & 0.333 & - & - \\
	AOL-NLM & 0.256 & 0.241 & 0.170 & 0.474 & - & - \\
	KnowHow-NLM & 0.166 & 0.147 & 0.108 & 0.575 & - & - \\
	WikiAnswers-NLM & 0.163 & 0.121 & 0.088 & 0.650 & - & - \\
	AOL-Seq2Seq & 0.283 & 0.181 & 0.156 & 0.765 & 0.043 & 0.031 \\
	KnowHow-Seq2Seq & 0.158 & 0.111 & 0.079 & 0.813 & 0.206 & 0.148\\
	\midrule
	Keyphrase-based~\cite{Garigliotti:2017:GQS}  
	 & 0.321 & 0.239 & 0.130 & - & 0.575 & 0.504\\
	Google API & 0.267 & 0.134 & 0.078 & - & 0.289 & 0.145 \\
	\bottomrule
\end{tabular}
\label{tbl:res_sg}
\end{table*}

We find that the AOL query log is the best source for generating QC suggestions, while KnowHow works well for QR.  AOL-Seq2Seq performs poorly on QR; this is because only 2\% of all training instances in $\langle \mathcal{Q},\mathcal{S} \rangle$ are QR pairs. 
For QC suggestions, the performance of AOL-PopSuffix and AOL-NLM are close to that of the Google API, while AOL-Seq2Seq even outperforms it. For QR suggestions, the performance of AOL-Seq2Seq is close to that of the Google API in terms of P@20, but is lower on P@10.  In general, our system is able to produce more suggestions than what the (public) Google API provides.  

Additionally, we also evaluate the recall of each QC method, using all relevant suggestions found as our recall base.
We further report on cumulative recall, i.e., for line $i$ of Table~\ref{tbl:res_sg} it is the recall of methods from lines $1..i$ combined together. 
We observe that each configuration brings a considerable improvement to cumulative recall. This shows that they generate unique query suggestions. 
For example, given the query ``choose bathroom,'' our system generates unique query suggestions from different methods and sources, e.g., ``choose bathroom marks'' (WikiAnswers-NLM), ``choose bathroom supply'' (AOL-NLM), ``choose bathroom for your children'' (KnowHow-NLM), ``choose bathroom grout'' and ``choose bathroom appliances'' (KnowHow-Seq2Seq), which are beyond what the Google API and the keyphrase-based system~\citep{Garigliotti:2017:GQS} provide. 
Therefore it is beneficial to combine suggestions from multiple configurations for further reranking.

\section{Conclusions}
\label{sec:conl}

We have addressed the task of generating query suggestions that can assist users in completing their tasks.  
We have focused on the first component of a two-step pipeline, dedicated to creating suggestion candidates.  For this, we have considered several methods and multiple information sources.  We have based our evaluation on the TREC Tasks track, and collected a large number of annotations via crowdsourcing.  Our results have shown that we are able to generate high-quality suggestion candidates.  We have further observed that the different methods and information sources lead to distinct candidates.

As our next step, we will focus on the second component of the pipeline, namely, suggestions ranking.    
As part of this component, we also plan to address the specific issue of subtasks coverage, i.e., improving the diversity of query suggestions.  

\FloatBarrier  

\renewcommand*{\bibfont}{\scriptsize}
\bibliographystyle{abbrvnat}
\bibliography{ecir2018-neuqs}

\end{document}